\documentclass[rsi,amsmath,amssymb,reprint,floatfix]{revtex4-1}

\usepackage[utf8]{inputenc}
\usepackage{graphicx}
\usepackage{dcolumn}
\usepackage{bm}
\usepackage{float}
\usepackage{footmisc}
\usepackage[utf8]{inputenc}
\usepackage{xcolor}
\usepackage[T1]{fontenc}
\usepackage{mathptmx}
\usepackage{xspace}
\usepackage{textcomp,gensymb}
\usepackage{hyperref}
\usepackage{siunitx}
\usepackage{ulem}
\DeclareSIUnit\gauss{G}

\usepackage[
disable
]{endfloat}
\begin{document}

\preprint{AIP/123-QED}

\title{A device for magnetic-field angle control in magneto-optical filters using a solenoid-permanent magnet pair}

\author{Sharaa A. Alqarni, Jack D. Briscoe, Clare R. Higgins, Fraser D. Logue, Danielle Pizzey,  Thomas G. Robertson-Brown \& Ifan G. Hughes}
\email{i.g.hughes@durham.ac.uk}
\affiliation{Department of Physics, Durham University, South Road, Durham, DH1 3LE, United Kingdom}

\date{\today}

\begin{abstract}

Atomic bandpass filters are used in a variety of applications due to their narrow bandwidths and high transmission at specific frequencies. Predominantly these filters in the Faraday (Voigt) geometry, using an applied axial(transverse) magnetic field with respect to the laser propagation direction. Recently, there has been interest in filters realized with arbitrary-angle magnetic fields, which have been made by rotating permanent magnets with respect to the $k$-vector of the interrogating laser beam. However, the magnetic-field angle achievable with this method is limited as field uniformity across the cell decreases as the rotation angle increases. In this work, we propose and demonstrate a new method of generating an arbitrary-angle magnetic field, using  a solenoid to produce a small, and easily alterable, axial field, in conjunction with fixed permanent magnets to produce a large transverse field. We directly measure the fields produced by both methods, finding them to be very similar over the length of the vapor cell. We then compare the transmission profiles of filters produced using both methods, again finding excellent agreement. Finally, we demonstrate the sensitivity of filter profile to changing magnetic-field angle (solenoid current), which becomes easier to exploit with the much improved angle control and precision offered by our new design.

\end{abstract}

\maketitle

\section{\label{sec:intro}Introduction}

Narrowband magneto-optical bandpass filters~\cite{dick1991ultrahigh, chen1993sodium,yeh1982dispersive,zielinska2012ultranarrow, Zentile2015g, Gerhardt:18, logue2022better} find great utility across a range of disciplines, including solar monitoring~\cite{Ohman1956b, Cimino1968, Cacciani1978, cacciani1990solar,  wrro175746}, atmospheric LIDAR~\cite{Fricke-Begemann:02, Popescu2004, Yang2011}, and  intra-cavity laser
frequency stabilization~\cite{Keaveney2016c, chang2022frequency,shi2020dual, liu2023voigt}. 
The spectrum of the light transmitted through an atomic vapor cell subject to an external magnetic field is dependent on the relative orientation of the magnetic field and the $k$-vector of the light. The most commonly used geometries are the Faraday configuration~\cite{faraday1846experimental,  harrell2009sodium, Kiefer2014}, where the magnetic field is parallel to the $k$-vector of the interrogating light, and Voigt configuration ~\cite{voigt1899theorie}, where the magnetic field is perpendicular to the $k$-vector~\cite{Voigt5, schuller1991voigt, Voigt1F, Voigt2F, Voigt3F, Voigt4, pyragius2019voigt, muroo1994resonant, briscoe2023voigt}. The general case with an arbitrary angle between the magnetic field and the axis of propagation is more difficult to treat mathematically -- and to optimize experimentally -- as the working angular range of the magneto-optical filter is limited and slight deviations from the optimum angle lead to reduced filter efficiency and spectral distortion. Consequently, there are far fewer experimental studies of this case owing to difficulties in setting and controlling the magnetic-field angle without encroaching the line-of-sight propagation~\cite{edwards1995magneto, nienhuis1998magneto, Rotondaro2015}. Nevertheless, there has been a recent burgeoning of interest in this geometry, as it offers the possibility of realizing better magneto-optical filters, when compared to Faraday and Voigt geometries~\cite{Rotondaro2015, Keaveney2018b}. Ongoing research and development efforts aim to address the experimental challenges and improve the performance, stability, and cost-effectiveness of these devices.

\begin{figure}[tbh!]
\begin{center}
\includegraphics[width=85mm,clip=true,trim = 0mm 1mm 0mm 0mm]{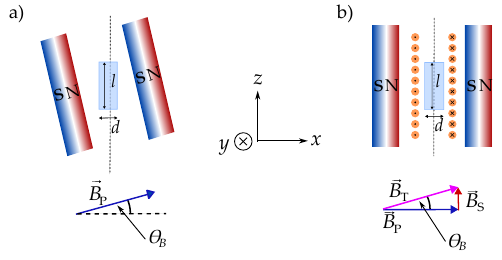}
\end{center}
\vspace{0mm}
\caption{An arbitrary magnetic-field angle with respect to the axis of the laser beam (i.e. along $z$), represented by the vertical, black, dashed line, can be produced by either: a) A Voigt magnetic field ($\vec{B_{\mathrm{P}}}$) set-up rotated by $\theta_{B}$ about the $y$-axis, as shown by the blue vector arrow or; b) A fixed Voigt magnetic field ($\vec{B_{\mathrm{P}}}$, blue arrow) and a tuneable solenoid magnetic field ($\vec{B_{\mathrm{S}}}$, red arrow) produce a combined magnetic field ($\vec{B_{\mathrm{T}}}$, pink arrow) at an angle of $\theta_{B}$ with respect to the $z$-axis. The red-blue rectangles represent a N-S permanent magnet, while the orange circles represent the solenoid, with either a dot or cross showing the current direction. The light blue rectangle shown in the center of the magnet arrangements represents a cylindrical atomic vapor cell of length $l$ and diameter $d$.}
\label{fig:comparison_setup}
\end{figure}

In our previous work involving arbitrary-angle filters \cite{Keaveney2018b} the magnetic field was controlled using a pair of permanent magnets positioned either side of the vapor cell, like those used in a Voigt geometry set-up, but rotated relative to the beam axis. The field strength was set by the magnet remanence field and separation, while the angle was set by physically rotating the magnets with respect to the $k$-vector of the laser beam; this concept is illustrated in Figure~\ref{fig:comparison_setup}a).

Maintaining magnetic field homogeneity at the 1\% level over millimeter vapor cell length scales is trivial~\cite{Danthesis}, but the use of these short cells comes at the expense of the requirement of elevated operating temperatures to produce sufficient atomic vapor density; this leads to self-broadening of spectral lines \cite{weller2011absolute} and ultimately reduced magneto-optical filter performance \cite{Zentile_2015}. With open-source magnetic field computation programmes \cite{ORTNER2020100466} becoming readily available, designing bespoke magnetic field profiles with field homogeneity extending tens of millimeters is now feasible \cite{pizzey2021tunable}, meaning standard `off-the-shelf' vapor cells -- with length of tens of millimeters -- can be used, with correspondingly lower operating temperatures required. However, this homogeneity along a given axis does not remain when the magnets are rotated, so the angular range of permanent magnet arbitrary-angle filter set-up is still limited, and the longer the cell, the greater the limitation.

To address these challenges, we suggest and implement an alternative approach to generate an arbitrary magnetic-field angle while maintaining the same field strength. Our approach involves incorporating an air-core solenoid between a pair of Voigt geometry permanent magnets generating a strong transverse magnetic field, with the vapor cell seated within the bore of the solenoid. A weak axial magnetic field is generated by the solenoid, and the magnitude of this field, and therefore the angle of the total field, can be regulated by controlling the current. This concept is illustrated in Figure~\ref{fig:comparison_setup}~b).

In this paper, we demonstrate that this combination of Voigt-geometry permanent magnets and a Faraday-geometry solenoid effectively addresses the experimental challenges associated with precise control of small magnetic-field angles, as well as removing the physical limit on achievable angle. Throughout the paper we will refer to this new method as `solenoid-plus-permanent', and the old method as `rotated-permanent'.

\par
The remainder of the paper is organized as follows: in Section~\ref{sec:Req_MOF} we explain the requirements of the magneto-optical filter and how we come to realize the parameters;  in Section~\ref{sec:design_overview} we present the magnetic field computation and compare the rotated-permanent and solenoid-plus-permanent configurations; in Section~\ref{sec:Bothresults} we present measurements of the magnetic field using a Hall probe for the two geometries and compare the filter performance of each; and finally conclusions are drawn and an outlook provided in Section~\ref{sec:conclusion}.

\section{\label{sec:Req_MOF}Requirements of the magneto-optical filter}

\begin{figure}[tbh!]
\begin{center}
\includegraphics[width=85mm,clip=true,trim = 0mm 0mm 0mm 0mm]{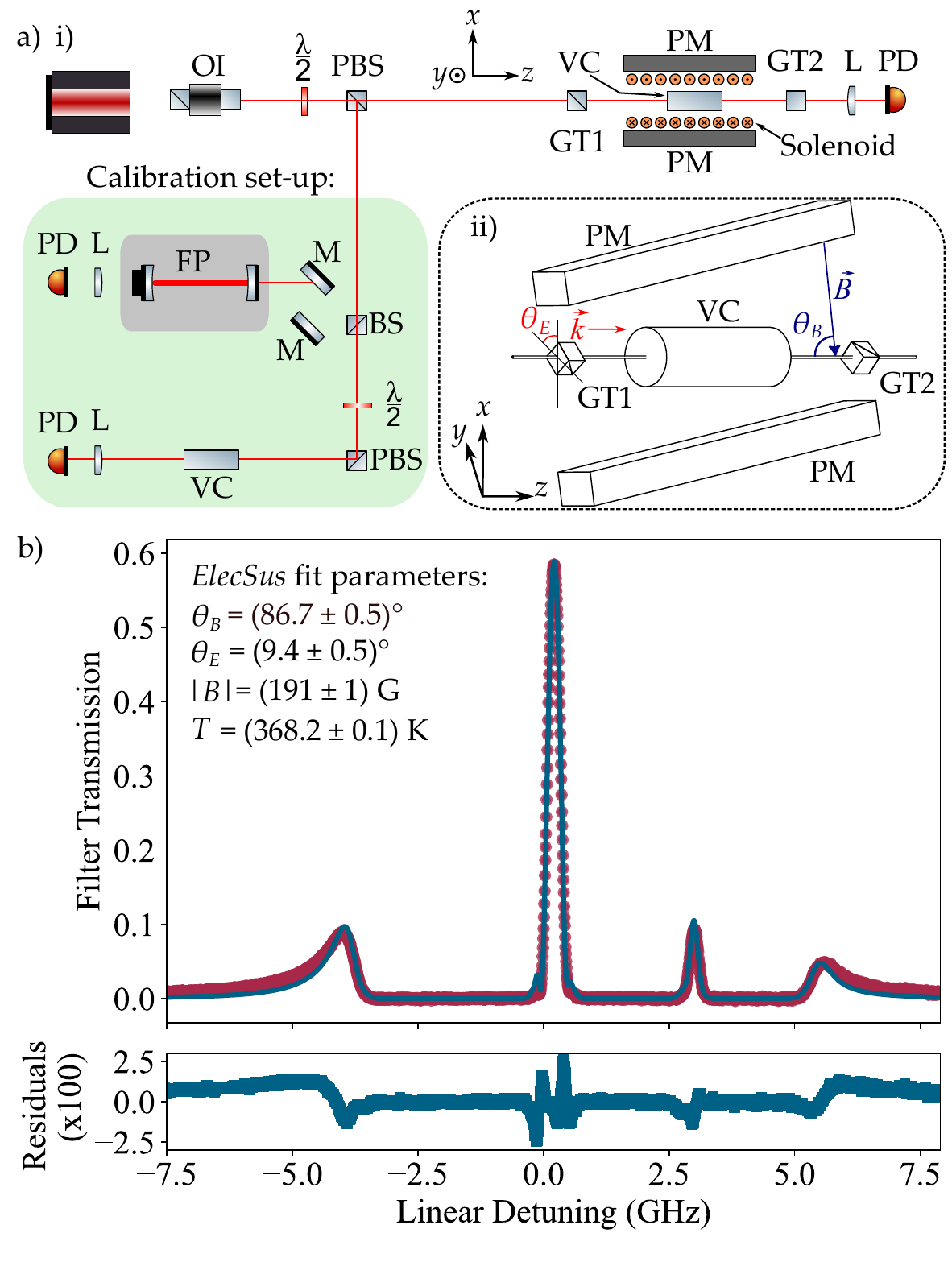}
\end{center}
\vspace{0mm}
\caption{a) i) Schematic of the solenoid-plus-permanent magneto-optical filter set-up and geometry. The filter consists of an atomic vapor cell in an applied magnetic field ($\vec{B}$) formed from two rectangular permanent magnets (PM) and a solenoid. The field strength is determined by the separation of the magnets and is adjustable up to 190\,G, while the magnetic-field angle $\theta_{B}$ is controlled by the current through the solenoid. Shown in the inset a) ii) is an illustration of the magnet set-up for the rotated-permanent set-up. In both methods the magnetic field is orientated in $x$-$z$-plane at an angle $\theta_{B}$ to the $z$-axis. An input high-extinction Glan-Taylor polarizer (GT1) is set at an angle $\theta_{E}$ with respect to the $x$-axis. The output polarizer (GT2) is crossed at 90$^{\circ}$ to the input polarizer. The transmission of the filter is measured via a lens (L) and photodetector (PD). Also shown is the calibration set-up, which includes a Fabry-Pérot etalon (FP) for linearizing the laser scan and a vapor cell (VC) for an absolute frequency reference~\cite{keaveney2014collective,pizzey2022laser}. b) Experimental transmission (blue data points) as a function of linear detuning of an arbitrary angle magneto-optical filter for a naturally abundant Rb vapor cell of $l$~=~75~mm. The magnetic-field angle $\theta_{B}$ was produced by rotating permanent magnets, as shown in a) ii). A theoretical $ElecSus$ fit (red solid line), with corresponding residuals, is shown. }
\label{fig:set-up}
\end{figure}

Figure~\ref{fig:set-up}~a) i) illustrates a schematic of the optical apparatus required for a Rb magneto-optical filter and the geometry of the set-up. Light emitted from an external cavity diode (ECD) laser with a center wavelength of 780 nm traverses an optical isolator (OI) and is divided into two separate beams using a polarizing beam-splitter (PBS) and a half-wave retarder plate ($\lambda$/2); one path goes to the magneto-optical filter, the other is for calibration of the frequency axis. Due to the non-linear response of the laser piezo that controls the output frequency, we need to calibrate the laser scan~\cite{harris2008optimization,pizzey2022laser} so that we can compare our transmission spectrum with theory. We follow the methods described in Pizzey \textit{et al.}~\cite{pizzey2022laser} and use a Fabry-Perot etalon for linearization and a reference 50 mm length natural abundance Rb cell for defining zero-detuning, which is chosen to be the weighted center of the line~\cite{Siddons2008b}.
\par
In the magneto-optical filter, the magnetic field vector, along the length of the vapor cell, is oriented in the $x$-$z$-plane at an angle of  $\theta_{B}$ to the $z$-axis, where $\theta_{B}$~=~\SI{0}{\degree} and $\theta_{B}$~=~\SI{90}{\degree} correspond to the Faraday and Voigt geometries, respectively. The vapor cell is positioned between two high-extinction polarizers (shown in Figure~\ref{fig:set-up}~a) ii)). The angle between the electric field vector of the light and the $x$-axis, $\theta_{E}$, influences the coupling between atomic transitions and polarization modes of the light. The angle of the input polarizer, $\theta_{E}$, can be adjusted, but the relative angle of the two polarizers GT1 and GT2 remains constant at \SI{90}{\degree} (i.e., crossed polarizers). This ensures that the transmission is zero in the absence of any atom-light interaction. 


\begin{figure*}[htbp]
\begin{center}
\includegraphics[width=\linewidth]{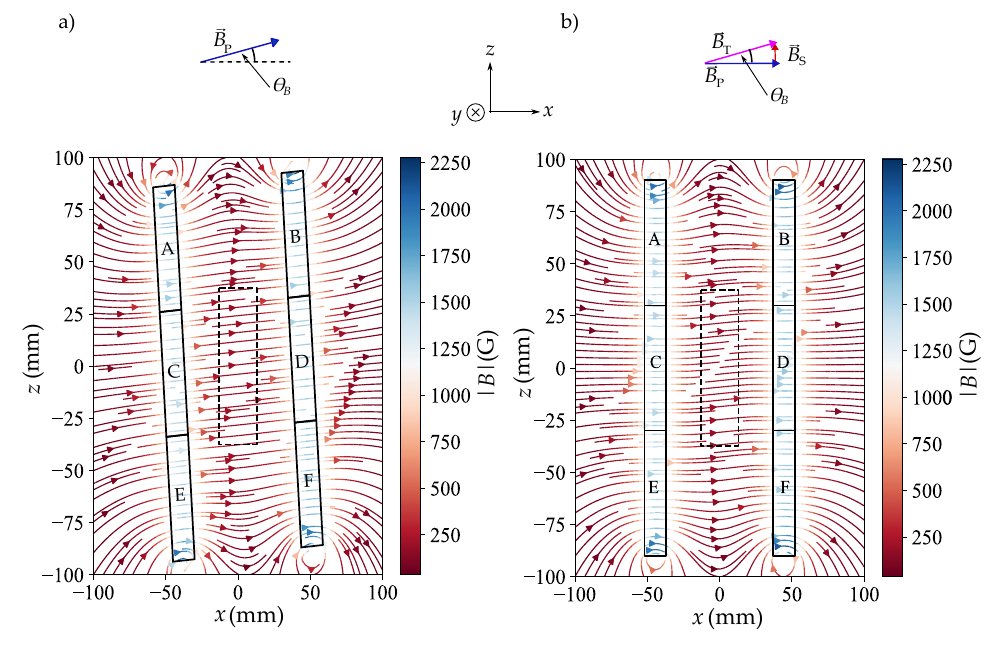}
\end{center}
\vspace{0mm}
\caption{$Magpylib$ magnetic field simulations for the magnetic field configurations shown in Figure~\ref{fig:comparison_setup} for six permanent magnets labelled A-F. The physical profile of each individual magnet is shown by a black solid line in the contour plots and the field strengths of each magnet, which vary by 5$\%$ between the weakest and strongest, are accounted for in the $Magpylib$ model. The arrows indicate the magnetic field vector's direction, while the color, and accompanying colorbar, represent the magnitude of the magnetic field. a) rotated-permanent configuration -- Voigt permanent magnets rotated by an angle $\theta_{B}$~=~4$^{\circ}$. b) solenoid-plus-permanent configuration -- Voigt configuration permanent magnets, and a solenoid current of 525 mA, which generates a $B$-field along $z$ of 13\,G. The physical profile of the solenoid wires are not displayed in the contour plot. } 
\label{fig:magnet_comp}
\end{figure*}

Using the open-source computer program $ElecSus$~\cite{zentile2015elecsus, keaveney2018elecsus}, we model the transmission of a weak laser beam~\cite{sherlock2009weak} through an alkali-metal atomic vapor with a given input polarization, magnetic-field strength and angle. We implement an extension to $ElecSus$ to calculate the transmission spectrum of the magneto-optical filter by calculating the subsequent transmission through a polarizer (crossed with respect to the input polarizer) after the vapor cell. A fitting routine can be implemented to optimize the filter peak transmission and linewidth by varying input parameters, such as the vapor temperature, magnetic-field strength and magnetic-field angle. 

In previous work~\cite{Keaveney2018b}, optimum magnetic-field strength, angle, and atomic vapor temperature were found for a Rb D2 line filter (natural abundance ratio) using a 5~mm length cell. In this work, however, the magneto-optical filter parameters were not optimized. Instead, the parameters were chosen to work robustly for a vapor cell of length 75~mm since the primary focus of this research is to conduct a comparative analysis between two methods of generating an arbitrary magnetic field, rather than fine-tuning the filter parameters for optimal performance. We used $ElecSus$ to identify suitable magnetic field parameters, which give a filter profile with a narrow peak and reasonable transmission at line center, for a filter using a $l$~=~75~mm vapor cell. We select a field magnitude of 190~G at an angle, $\theta_{B}$, of 86$^{\circ}$; these parameters also fall within the allowable angle constraints of the rotated-permanent configuration. 

We construct a filter with these parameters using the rotated-permanent set-up, the normalized transmission of which is shown in Figure~\ref{fig:set-up}~b). Here, the atomic vapor cell temperature is 368~K. Experimental data are displayed as red points, and an $ElecSus$ fit to the data is shown by a solid blue line, with fit parameters as displayed in the figure. Residuals are plotted underneath, showing very good agreement between theory and experiment. As previously demonstrated \cite{Keaveney2018b,Higgins:20}, we see that the $ElecSus$ model describes the magneto-optical filter behaviour well, and we will use this later in the work to further test the effectiveness of our new arbitrary-angle-field generation method.

\section{\label{sec:design_overview}Arbitrary-angle magnetic field control}

We will simulate the magnetic field profiles of two designs -- rotated-permanent (Figure~\ref{fig:comparison_setup}\,a)) and  solenoid-plus-permanent (Figure~\ref{fig:comparison_setup}\,b)) -- and compare the field homogeneity and tolerances. For the solenoid-plus-permanent configuration to produce a magnetic-field angle of $\theta_{B}$~=~86$^\circ$, we require the solenoid to produce an axial magnetic field of 13\,G over a length scale of $l$~=~75~mm.

We use $Magpylib$~\cite{ORTNER2020100466}, an open-source python package for magnetic field computation, to simulate the magnetic field geometry. To produce the field strength required for the magneto-optical filter discussed in Section~\ref{sec:Req_MOF}, we employed ``off-the-shelf'' commercially available strontium ferrite permanent magnets of grade Y30BH, due to their easy availability. These magnets are cuboidal, measuring 15~mm~x~20~mm~x~60 mm along the $x$, $y$, and $z$ axes, respectively. They are magnetized along the $x$-axis, with each individual magnet having a slightly different strength, with a variation of 5\% between the weakest and strongest. To ensure field homogeneity across the 75 mm length of the vapor cell, we stacked three magnets along the $z$-axis for each half of the set-up. This arrangement resulted in a total of six magnets forming the Voigt permanent magnet configuration. 3D printed plastic holders were used, each holding three magnets. A pair of symmetrical holders constituted the Voigt permanent magnet geometry, with the separation of magnets along the $x$-axis being adjustable. The field strength is determined by the distance between the permanent magnets, which we set to be 74 mm, giving a field strength of |$B$|~=~190~G. The Voigt permanent magnet configuration is mounted on a Thorlabs rotating breadboard featuring a removable center portion (RBB300A/M). This set-up enables the vapor cell to remain fixed with respect to the laser beam axis (i.e. $z$-axis) while allowing for the rotation of the Voigt magnets to generate the desired arbitrary magnetic-field angle. Figure~\ref{fig:magnet_comp}~a)~illustrates the $Magpylib$ simulations of the rotated-permanent configuration. The arrows indicate the direction of the magnetic field vectors, while the color and accompanying colorbar illustrate the strength of the field. The black dashed line between the Voigt magnets illustrate the physical profile of the $l$~=~75~mm vapor cell.

\begin{figure}[tbh!]
\begin{center}
\includegraphics[width=\linewidth]{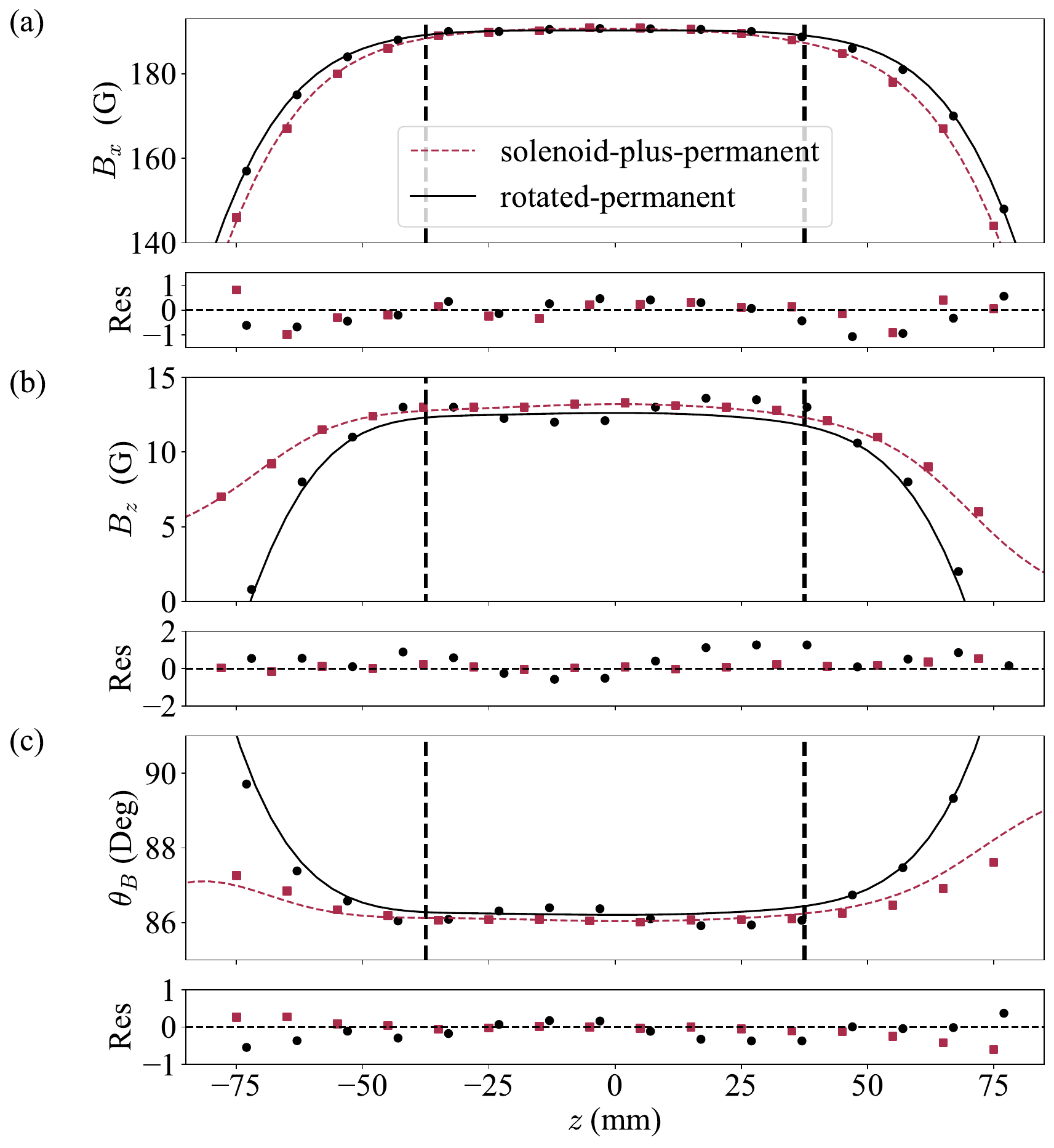}
\end{center}
\vspace{0mm}
\caption{Comparison of measured magnetic field components:  (a) The transverse magnetic field ($B_x$) is plotted for the two different methods: solenoid-plus-permanent (red squares), and rotated-permanent (black circles). The theoretical model for each configuration is shown as a black solid line (rotated-permanent geometry), and a red dashed line (solenoid-plus-permanent). The two methods produce similar, uniform fields of \SI{190}{\gauss} over the position of the vapor cell, which is indicated by vertical black dashed lines. (b) The axial magnetic field ($B_z$) is plotted for the same configurations and has a value of \SI{13}{\gauss} over the length of the vapor cell. (c) $\theta_{B}$, the angle of the magnetic field vector with respect to the $z$-axis. This is calculated from the measurements in a) and b). In both configurations, $\theta_{B}$ is approximately \SI{86}{\degree}. Residuals are shown below each subplot, illustrating an excellent agreement between the measured data and the theoretical model predictions \cite{hughes2010measurements}.}
\label{fig:magnetic field and angle}
\end{figure}

The solenoid is required to be longer than the length of the vapor cell for magnetic field homogeneity, and possess a central bore capable of accommodating both the vapor cell and its heater. Additionally, the solenoid should generate an axial magnetic field of 13~G without requiring excessive current to prevent overheating of the solenoid wires. According to $Magpylib$ simulations, we established that a solenoid with a length of 140~mm and an inner diameter (ID) of 45~mm, consisting of two layers of 156 turns of wire with a thickness of 0.9 mm, would yield the desired axial field when supplied with less than 1 A of current. The solenoid is formed by winding copper wire around a cylindrical PTFE former of the appropriate dimensions, ensuring thermal isolation between the vapor cell and the solenoid; this allows us to have independent control over the magnetic field and the temperature of the vapor cell. The solenoid is also mounted within the center portion of the rotating breadboard, such that when the Voigt permanent magnets are rotated, the solenoid and vapor cell remain stationary. Figure~\ref{fig:magnet_comp}~b)~illustrates the $Magpylib$ simulations of the solenoid-plus-permanent configuration, with a solenoid current of 525\,mA. It can be seen that along the $z$-axis, within the $z$-range of the vapor cell (dashed black lines) the field magnitude and direction of the two configurations are almost identical. This is shown explicitly in the theory lines of Figure~\ref{fig:magnetic field and angle}. 

We also use $Magpylib$ to simulate the effect of changing angle on the uniformity of the magnetic field over the extent of the vapor cell for the two methods. We see that $|B|$ is much more uniform over the extent of the vapor cell using the solenoid-plus-permanent method; for example, at $\theta_B =$~\SI{70}{\degree} the solenoid-plus-permanent configuration has a range of \SI{0.3}{\gauss} and rotated-permanent has range of \SI{4}{\gauss}.

\section{\label{sec:Bothresults}Results}

We compare the two methods of generating the arbitrary-angle magnetic field through two approaches: first, by measuring the axial and transverse magnetic fields using a Hall probe; and second, by utilizing the atoms as magnetic field sensors within the magneto-optical filter.

\subsection{\label{sec:hall_probe_results}Hall probe measurements}
A transverse Hall probe (Magnetic Instruments GM08 Gaussmeter) was employed to measure the transverse field component, $B_x$, along the axis of the laser beam ($z$-axis). Figure~\ref{fig:magnetic field and angle}\,a) shows two experimentally measured field profiles (data points) as a function of $z$: the rotated-permanent geometry (black), and the solenoid-plus-permanent (red). Theoretical field profiles, calculated with $Magpylib$, are represented by solid or dashed lines in the corresponding colors. Notably, the maximum measured transverse magnetic field of two configurations was the same, at 190~G, and field homogeneity is maintained at 1.5\% over the region occupied by the vapor cell, indicated by the dashed black lines. The residuals show excellent agreement between the data and $Magpylib$ model.

The axial magnetic field component $B_z$ was measured using an axial Hall probe, and the results are depicted in Figure~\ref{fig:magnetic field and angle}\,b). Results show good agreement between the two methods. For both, the maximum measured value of the field was 13 G, as expected. Residuals show excellent agreement between the experimental measurements and theoretical predictions. Again, the uniformity of the magnetic field along the length of the vapor cell was confirmed. 

The magnetic-field angle, $\theta_{B}$, for the two configurations was calculated from the $B_x$ and $B_z$ using $\tan ^{-1} (B_z/B_x)$, and is depicted in Figure~\ref{fig:magnetic field and angle}\,c).  {$\theta_{B}$ is approximately 86$^\circ$ for both geometries, with a RMS error of \SI{0.2}{\degree} between the two methods.}

\subsection{\label{sec:results}Magneto-optical filter measurements}

The results presented in Section~\ref{sec:hall_probe_results} indicate minimal discrepancies in the magnetic-field strength and angle between the two magnetic field configurations. Therefore we expect magneto-optical filter transmission profiles realized with the two configurations to be very similar.  In this section, we construct magneto-optical filters with both field configurations, and compare their profiles. In addition, we tune the angle of the magnetic field by changing the current in the solenoid.

\begin{figure}[tbh!]
\begin{center}
\includegraphics[width=\linewidth]{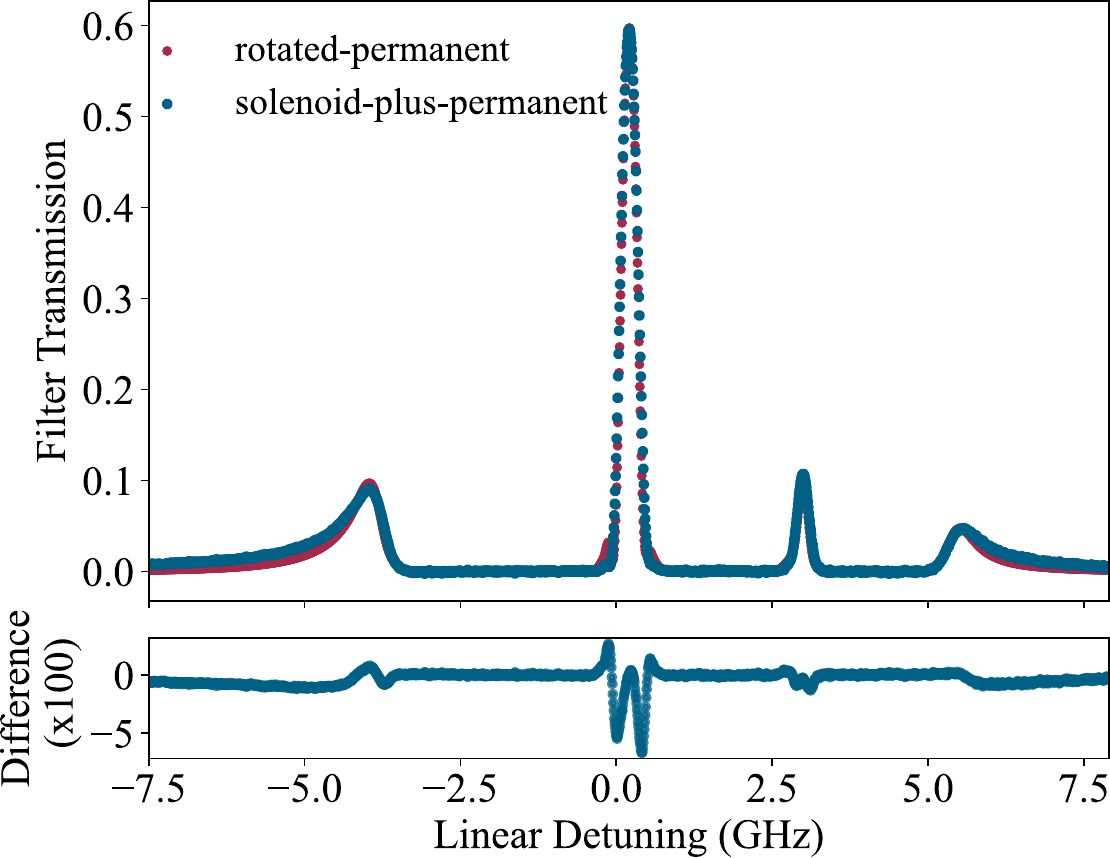}
\end{center}
\vspace{0mm}
\caption{Comparison between rotated-permanent (red points) and solenoid-plus-permanent (blue points) filter profiles. Both are for a natural abundance Rb D2 transitions through a 75 mm vapor cell in the weak probe regime as a function of linear detuning. 
Plotted underneath are the difference between the two experiments which shows excellent agreement.}

\label{fig:rotated magnets and solenoid comarsion}
\end{figure}

Figure~\ref{fig:rotated magnets and solenoid comarsion} compares the performance of the arbitrary-angle magnetic-field filter produced by the rotated-permanent geometry with that produced using a solenoid-plus-permanent set-up. The difference between the two profiles is shown in the bottom subplot. We see that there is excellent agreement between the two methods of producing the arbitrary magnetic-field angle; all the features of the profile are reproduced.

\begin{figure*}[tbh!]
\begin{center}
\includegraphics[width=\linewidth]{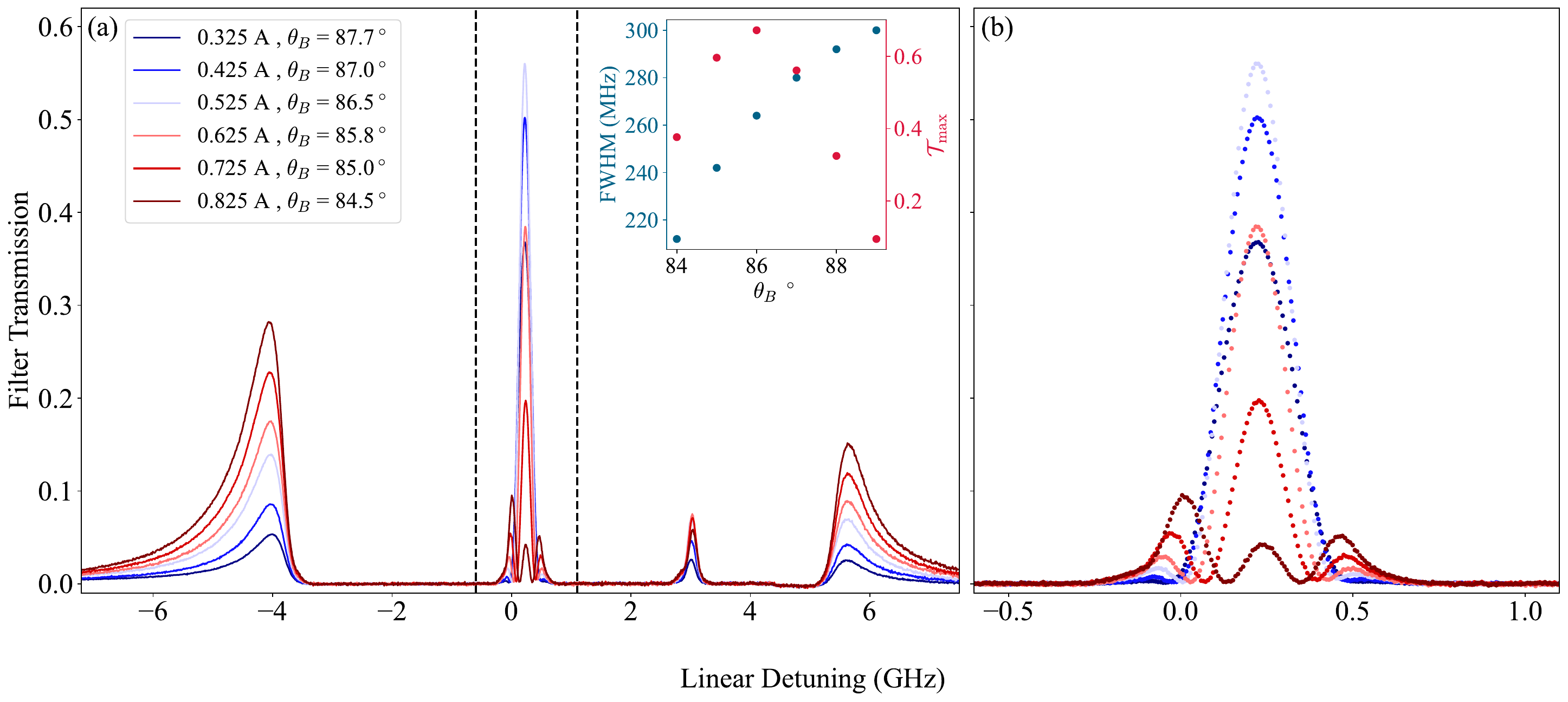}
\end{center}
\vspace{0mm}
\caption{a) Experimental Rb D2 line magneto-optical filter transmission, through a natural abundance Rb vapor cell of length $l$ = 75 mm, as a function of linear detuning in the weak probe regime. The plot shows the effect of changing solenoid current, and correspondingly the angle, $\theta_B$,  of the total magnetic field, on the filter spectra with the solenoid-plus-permanent configuration. Quoted angles are extracted from $ElecSus$ fits. b) shows an expanded view of the central peaks of the spectra. The inset shows theoretical predictions from $ElecSus$ of the behaviour of the full-width-at-half-maximum (FWHM) and the maximum transmission of the central filter peak as magnetic-field angle is varied. 
}
\label{fig:filter at dif currents}
\end{figure*}

The solenoid offers an additional benefit by providing us with more efficient and rapid tuning capabilities. In contrast to the rotated-permanent configuration, where rotating the magnets to generate the arbitrary-angle magnetic field can be a slow process, we can swiftly adjust the magnetic-field angle by simply changing the current supplied to the solenoid. The solenoid method also makes precisely selecting the angle easier, as this fine tuning is difficult when rotating permanent magnets by hand. This enhances the versatility and responsiveness of our experimental set-up.

Figure~\ref{fig:filter at dif currents}\,a) demonstrates the diverse filter spectra generated using the solenoid-plus-permanent configuration with varying current, as well as the effect on filter characteristics. The dark blue trace shows the filter spectrum when a current of 0.325\,A is applied to the solenoid. Increasing the solenoid current increases not only the height of the central peak, but also the height of the other peaks. However, once the current reaches 0.625\,A (i.e. $\theta_{B}$~=~85$^\circ$), an interesting change occurs: the transmission of the central peak starts to decrease, while the peaks at the wings continue to rise. A zoom in to the central region of the filters, as shown in Figure~\ref{fig:filter at dif currents} b), highlights the response of the main peak to varying currents. We also note that as the current increases, the main peak width increases continually. This central peak height and width behaviour with changing angle, $\theta_{B}$, are shown in the inset in Figure~\ref{fig:filter at dif currents}. 

The sensitivity of the magneto-optical filter response to a small angle change further reinforces why the solenoid-plus-permanent configuration is better suited for optimizing magneto-optical filters, compared to using the rotated-permanent approach, since the angle can be precisely controlled. Indeed, the solenoid-plus-permanent configuration has recently been used to demonstrate the magneto-optical filter with the highest recorded figure of merit to date~\cite{Fraser-arxiv}. Figure~6~b) demonstrates that a large change in peak transmission can be achieved by a small angular change of the direction of the magnetic field; this has the potential to be the basis of an optical switch \cite{raevsky1999stabilizing,  jelinkova2004linbo3}.

The largest angle solenoid-plus-permanent filter created and analysed here had $\theta_B$ of \SI{84.5}{\degree}. However the equipment used in this work has been used to produce an angle of $\theta_B$ = \SI{66}{\degree}. This value is limited by the chosen solenoid characteristics, and the power supply. Larger $B_z$, and correspondingly smaller $\theta_B$ could easily be produced using a different solenoid/power supply combination. Indeed, solenoids have been used to generate fields exceeding 4\,kG in magneto-optical filter experiments, though this requires water cooling \cite{Kiefer2014}. It would therefore be possible to create a solenoid-permanent magnet set-up capable of producing any chosen field orientation. It should be noted, however, that in the case of large magnetic-field angles using this set-up, the resultant field magnitude is highly dependent on the field angle. 

This wide range of achievable angles is in contrast to the rotated-permanent configuration, which has a physical limit (for our chosen magnets) of \SI{70}{\degree}, though the field-non-uniformity over the cell is a limiting factor well before this angle is reached.

\section{\label{sec:conclusion}Conclusion}

In this study, we proposed a new method for generating an arbitrary magnetic-field angle, combining a fixed Voigt geometry permanent magnet pair with a solenoid. This method allows for more precise and flexible control of the magnetic-field angle than the previously used method of rotating the permanent magnet pair. We simulated the fields produced by the two methods with $Magpylib$, to select appropriate solenoid parameters to replicate the magnetic field created using the old method, with our new set-up. We then conducted an experimental comparative analysis of the fields produced by the two methods, finding excellent agreement with comparable measured field strengths and magneto-optical filter profiles. 

The main limitation of the rotated-permanent method is the non-uniformity of the field produced over the length of the cell at larger rotation angles. This problem is resolved in the solenoid-plus-permanent configuration because the permanent magnets remain fixed relative to the cell, and the relative uniformity of the solenoid is independent of the field produced.  Therefore larger angles can be created, and longer vapor cells can be used with this method, which will allow for better magneto-optical filters to be realized~\cite{Keaveney2018b}. We also show how small changes in magnetic-field angle can create vastly different filter profiles, which the precise angle control and flexibility of our new design make easier to exploit.

\section{\label{sec:ack}Acknowledgments}
We gratefully acknowledge Steven Wrathmall and Liam Gallagher for helpful discussions and EPSRC (Grant No. EP/R002061/1) and the UK Space Agency (Grant No. UKSAG22\_0031\_ETP2-035) for funding. Sharaa Alqarni acknowledges Najran University, Najran, KSA, for financial support.

\section{\label{sec:dec}Author Declarations}
The authors have no conflicts of interest to disclose.

\section{\label{sec:app}Data Availability}
The data that support the findings of this study are freely available in DRO at https://doi.org/10.15128/r2rb68xb893.

\bibliography{REFERENCES}
\end{document}